\documentstyle [12pt,amsfonts] {article}
\input epsf

\newcommand{\beq}{\begin{equation}}
\newcommand{\eeq}{\end{equation}}
\newcommand{\beqs}{\begin{eqnarray}}
\newcommand{\eeqs}{\end{eqnarray}}
\newcommand{\lsim}{\mathrel{\raisebox{-.6ex}{$\stackrel{\textstyle<}{\sim}$}}}

\catcode`@=11
\@addtoreset{equation}{section}
\@addtoreset{equation}{subsection}
\def\theequation{\ifnum\value{section}=0 \arabic{equation}\ignorespaces
\else \ifnum\value{section}=-1 A.\arabic{equation}\ignorespaces
\else \ifnum\value{subsection}=0 \thesection.\arabic{equation}\ignorespaces
\else \thesection.\arabic{subsection}.\arabic{equation}\ignorespaces
                           \fi
                      \fi
                 \fi}
\catcode`@=12

\begin{document}

\def\thefootnote{\fnsymbol{footnote}}

\baselineskip 6.0mm

\vspace{4mm}

\begin{center}

{\Large \bf Potts Model Partition Functions for Self-Dual Families of Strip
Graphs}

\vspace{8mm}

\setcounter{footnote}{0}
Shu-Chiuan Chang\footnote{email: shu-chiuan.chang@sunysb.edu} and
\setcounter{footnote}{6}
Robert Shrock\footnote{email: robert.shrock@sunysb.edu}

\vspace{6mm}

C. N. Yang Institute for Theoretical Physics  \\
State University of New York       \\
Stony Brook, N. Y. 11794-3840  \\

\vspace{10mm}

{\bf Abstract}
\end{center}

We consider the $q$-state Potts model on families of self-dual strip graphs
$G_D$ of the square lattice of width $L_y$ and arbitrarily great length $L_x$,
with periodic longitudinal boundary conditions. The general partition function
$Z$ and the $T=0$ antiferromagnetic special case $P$ (chromatic polynomial)
have the respective forms $\sum_{j=1}^{N_{F,L_y,\lambda}} c_{F,L_y,j}
(\lambda_{F,L_y,j})^{L_x}$, with $F=Z,P$.  For arbitrary $L_y$, we determine
(i) the general coefficient $c_{F,L_y,j}$ in terms of Chebyshev polynomials,
(ii) the number $n_F(L_y,d)$ of terms with each type of coefficient, and (iii)
the total number of terms $N_{F,L_y,\lambda}$.  We point out interesting
connections between the $n_Z(L_y,d)$ and Temperley-Lieb algebras, and between
the $N_{F,L_y,\lambda}$ and enumerations of directed lattice animals. Exact
calculations of $P$ are presented for $2 \le L_y \le 4$.  In the limit of
infinite length, we calculate the ground state degeneracy per site (exponent of
the ground state entropy), $W(q)$.  Generalizing $q$ from ${\mathbb Z}_+$ to
${\mathbb C}$, we determine the continuous locus ${\cal B}$ in the complex $q$
plane where $W(q)$ is singular.  We find the interesting result that for all
$L_y$ values considered, the maximal point at which ${\cal B}$ crosses the real
$q$ axis, denoted $q_c$ is the same, and is equal to the value for the infinite
square lattice, $q_c=3$. This is the first family of strip graphs of which we
are aware that exhibits this type of universality of $q_c$.

\vspace{16mm}

\pagestyle{empty}
\newpage

\pagestyle{plain}
\pagenumbering{arabic}
\renewcommand{\thefootnote}{\arabic{footnote}}
\setcounter{footnote}{0}

\section{Introduction}

The $q$-state Potts antiferromagnet (AF) \cite{potts,wurev} exhibits nonzero
ground state entropy, $S_0 > 0$ (without frustration) for sufficiently large
$q$ on a given lattice $\Lambda$ or, more generally, on a graph $G=(V,E)$
defined by its set of vertices $V$ and edges joining these vertices $E$. This
is equivalent to a ground state degeneracy per site $W > 1$, since $S_0 = k_B
\ln W$.  There is a close connection with graph theory here, since the
zero-temperature partition function of the above-mentioned $q$-state Potts
antiferromagnet on a graph $G$ satisfies
\beq
Z(G,q,T=0)_{PAF}=P(G,q)
\label{zp}
\eeq
where $P(G,q)$ is the chromatic polynomial expressing the number of ways of
coloring the vertices of the graph $G$ with $q$ colors such that no two
adjacent vertices have the same color (for reviews, see
\cite{rrev}-\cite{bbook}).  The minimum number of colors necessary for such a
coloring of $G$ is called the chromatic number, $\chi(G)$.  Thus
\beq
W(\{G\},q) = \lim_{n \to \infty} P(G,q)^{1/n}
\label{w}
\eeq
where $n=|V|$ is the number of vertices of $G$ and $\{G\} = \lim_{n \to
\infty}G$.  At certain special
points $q_s$ (typically $q_s=0,1,.., \chi(G)$), one has the noncommutativity of
limits
\beq
\lim_{q \to q_s} \lim_{n \to \infty} P(G,q)^{1/n} \ne \lim_{n \to
\infty} \lim_{q \to q_s}P(G,q)^{1/n}
\label{wnoncom}
\eeq
and hence it is necessary to specify the
order of the limits in the definition of $W(\{G\},q_s)$ \cite{w}. Denoting
$W_{qn}$ and $W_{nq}$ as the functions defined by the different order of limits
on the left and right-hand sides of (\ref{wnoncom}), we take $W \equiv W_{qn}$
here; this has the advantage of removing certain isolated discontinuities that
are present in $W_{nq}$.

Using the expression for $P(G,q)$, one can generalize $q$ from ${\mathbb Z}_+$
to ${\mathbb C}$.  The zeros of $P(G,q)$ in the complex $q$ plane are called
chromatic zeros; a subset of these may form an accumulation set in the $n \to
\infty$ limit, denoted ${\cal B}$ \cite{bkw}, which is the continuous locus of
points where $W(\{G\},q)$ is nonanalytic. (For some families of graphs 
${\cal B}$ may be null, and $W$ may also be nonanalytic at certain discrete 
points; this is not relevant for the present paper.)  The maximal region in
the complex $q$ plane to which one can analytically continue the function
$W(\{G\},q)$ from physical values where there is nonzero ground state entropy
is denoted $R_1$.  The maximal value of $q$ where ${\cal B}$ intersects the
(positive) real axis is labelled $q_c(\{G\})$.  This point is important since
it separates the interval of physical $q \ge q_c(\{G\})$ on the positive real
$q$ axis where the Potts model exhibits nonzero ground state entropy (without
frustration) from the interval $0 \le q \le q_c(\{G\})$ in which $W$ has
different analytic form(s).

\begin{figure}[h]
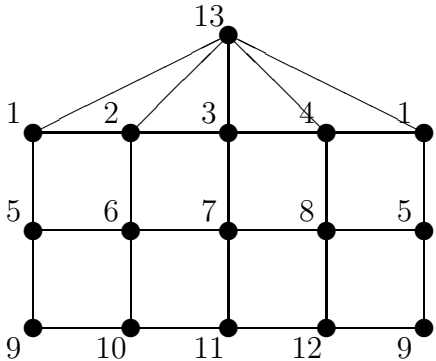

\caption{\footnotesize{Example of a strip graph $G_D( L_y \times L_x)$ for the
case $L_y=3$, $L_x=4$.}} 
\label{strip}
\end{figure}

In this paper we shall present a number of exact results on the Potts model
partition function and the special case comprised by the zero-temperature
Potts antiferromagnet for families of self-dual strip graphs of the
square lattice with (i) a fixed transverse width $L_y$, (ii) arbitrarily great
length $L_x$, (iii) periodic longitudinal boundary conditions, and (iv) such
that each vertex on one side of the strip, which we take to be the upper side
(with the strip oriented so that the longitudinal, $x$ direction is horizontal)
are joined by edges to a single external vertex.  A strip graph of this type
will be denoted generically as $G_D$ (where the subscript $D$ refers to the
self-duality) and, when its size is indicated, as $G_D(L_y \times L_x)$.  We
shall present a number of structural formulas that hold for arbitrary $L_y$.
After recalling our earlier work for the case $L_y=1$ \cite{w,wc}, we shall
present exact results for the chromatic polynomial for $L_y=2$ through $L_y=4$
and shall study the zeros of these polynomials and determine the
continuous accumulation set ${\cal B}$ of these zeros in the limit $L_x \to
\infty$.  In Fig. \ref{strip} we show an illustrative example of this family of
strip graphs, for the case $L_x=4$ and $L_y=3$.

\vspace{8mm}

\unitlength 1.3mm
\begin{picture}(40,30)
\multiput(30,0)(10,0){5}{\circle*{2}}
\multiput(30,10)(10,0){5}{\circle*{2}}
\multiput(30,20)(10,0){5}{\circle*{2}}
\multiput(30,0)(10,0){5}{\line(0,1){20}}
\multiput(30,0)(0,10){3}{\line(1,0){40}}
\put(50,30){\circle*{2}}
\put(30,20){\line(2,1){20}}
\put(40,20){\line(1,1){10}}
\put(50,20){\line(0,1){10}}
\put(60,20){\line(-1,1){10}}
\put(70,20){\line(-2,1){20}}
\put(28,-2){\makebox(0,0){9}}
\put(38,-2){\makebox(0,0){10}}
\put(48,-2){\makebox(0,0){11}}
\put(58,-2){\makebox(0,0){12}}
\put(68,-2){\makebox(0,0){9}}
\put(28,12){\makebox(0,0){5}}
\put(38,12){\makebox(0,0){6}}
\put(48,12){\makebox(0,0){7}}
\put(58,12){\makebox(0,0){8}}
\put(68,12){\makebox(0,0){5}}
\put(28,22){\makebox(0,0){1}}
\put(38,22){\makebox(0,0){2}}
\put(48,22){\makebox(0,0){3}}
\put(58,22){\makebox(0,0){4}}
\put(68,22){\makebox(0,0){1}}
\put(48,32){\makebox(0,0){13}}
\end{picture}

\vspace{8mm}

In general, the graph $G_D(L_y \times L_x)$ has 
\beq
n \equiv |V|= L_xL_y+1 = f
\label{n}
\eeq
vertices, equal to the number of faces, $f$, and 
\beq
e \equiv |E|=2L_xL_y
\label{e}
\eeq
edges. We recall that the dual $G^*$ of a planar graph $G=(V,E)$ is defined as
the graph obtained by replacing each vertex (face) of $G$ by a face (vertex) of
$G^*$ and connecting the vertices of the resultant $G^*$ by edges (bonds).  The
graph $G$ is self-dual if and only if $G= G^*$. It is easily checked that the
family $G_D(L_y \times L_x)$ is self-dual.  We recall that the degree $\Delta$
of a vertex is defined as the number of vertices to which it is adjacent (i.e.,
the number of nearest neighbors).  In the graph $G_D(L_y \times L_x)$ there are
$L_x(L_y-1)$ vertices with degree $\Delta=4$, $L_x$ vertices with $\Delta=3$,
and finally the one external vertex with $\Delta=L_x$.  The chromatic number is
\beq
\chi[G_D(L_y \times L_x)]=\chi[(Wh)_{L_x+1}] = \cases{ 3 & if $L_x$ is even \cr
                                                    4 & if $L_x$ is odd  \cr}
 \ . 
\label{chily1}
\eeq

The (infinite) square lattice (interpreted as the infinite limit of a graph) is
self-dual, and this property has been useful in studies of the Ising, and
general $q$-state Potts model on this lattice.  An early example of this was
the calculation by Kramers and Wannier of the critical temperature of the
(zero-field) Ising model on the square lattice \cite{kw} before Onsager derived
a closed-form expression for the free energy \cite{ons}.  The Potts model
partition function and the equivalent Tutte polynomial \cite{tutte2,tutte3} of
a planar graph $G$ satisfy certain symmetry relations when one changes $G$ to
$G^*$ \cite{wurev,a}.  Let us define $v=e^K-1$, where $K=J/(k_BT)$, with $T$
being the temperature and $J$ the spin-spin exchange constant for the Potts
model. Then \cite{wurev}
\beq
Z(G,q,v) = v^{e(G)}q^{-c(G)} Z(G^*,q,\frac{q}{v}) \ . 
\label{zdual}
\eeq
If $G=G^*$, this reduces to an identity relating the $q$-state Potts model on
$G$ to the same model with $v \to q/v$. In previous studies of
complex-temperature (Fisher) zeros of the Potts model, it has proved useful to
employ families of sections of the square lattice that are self-dual so that
one can take advantage of the symmetry property (\ref{zdual})
\cite{martin}-\cite{kc}.  The resulting patterns of zeros can be compared with
those obtained with other types of boundary conditions \cite{martin,ks,css}.
It may be recalled that in finite-lattice studies one often chooses periodic
boundary conditions to minimize boundary effects, but these do not preserve the
self-duality property of the infinite square lattice.  There are several types
of self-dual boundary conditions for strips of the square lattice
\cite{martin,wuetal,pfef}.  (The boundary conditions involved in the family
$G_D(L_y \times L_x)$ were denoted DBC2 in \cite{pfef}.)  In this paper,
instead of studying complex-temperature zeros, we shall focus on zeros of
$P(G,q)$ in the complex $q$ plane.  As we shall show, these exhibit some very
interesting features.  In particular, we shall show that for all of the strip
widths considered, we find a universal value of $q_c$, namely $q_c=3$, the
value for the infinite square lattice \cite{lenard}.

\section{Some General Properties} 

For a recursive family of graphs, such as the strip graphs considered in this
paper, comprised of $L_x$ repetitions of a basic subgraph, the $q$-state 
Potts model partition function for arbitrary $q$ (not necessarily in ${\mathbb
Z}_+$) and $v$ has the form \cite{a}
\beq
Z(G,q,v) = \sum_{j=1}^{N_{Z,G,\lambda}} c_{Z,G,j}(\lambda_{Z,G,j})^{L_x} 
\label{zgsum}
\eeq
where the coefficients $c_{Z,G,j}$ and the terms $\lambda_{Z,G,j}$ depend on
the lattice type, the boundary conditions, and the width, but not the length.
Since we are only dealing in this paper with a particular type of strip graph,
$G_D$, we shall henceforth usually suppress the $G_D$ in the notation, but
shall make explicit the dependence on $L_y$, writing $N_{Z,L_y,\lambda}$ for
the total number of terms, $\lambda_{Z,L_y,j}$ for a given term, $c_{Z,L_y,j}$
for the corresponding coefficient, and similarly for the chromatic polynomial.
Since the chromatic polynomial is a special case of the Potts model partition
function, it also has this form \cite{bkw}
\beq
P(G_D(L_y \times L_x),q) =  
\sum_{j=1}^{N_{P,L_y,\lambda}} c_{P,L_y,j}(\lambda_{P,L_y,j})^{L_x} \ . 
\label{pgsum}
\eeq 
Following our earlier nomenclature \cite{w}, we denote a $\lambda$ as leading 
(= dominant) if it has a magnitude greater than or equal to the magnitude of
other $\lambda$'s.  In the limit $n \to \infty$ the leading $\lambda$ in $Z$
determines the free energy per site $f=\lim_{n \to \infty} Z^{1/n}$ and
similarly the leading $\lambda$ in $P$ determines the function $W$ defined in
(\ref{w}).  The continuous locus ${\cal B}$ where $f$ or $W$ is nonanalytic
thus occurs where there is a switching of dominant $\lambda$'s in $Z$ and $P$,
respectively, and is the solution of the equation of degeneracy in magnitude of
these dominant $\lambda$'s.

In general one can regard the $\lambda$'s as the (nonzero) eigenvalues of a
certain coloring matrix \cite{b,matmeth,cf}.  In our earlier calculations, we
have found examples of zero eigenvalues \cite{cf}, but for the present family
we only obtain nonzero eigenvalues.  We note that while the coefficients
$c_{P,L_y,j}$ and $c_{Z,L_y,j}$ can be obtained as multiplicities of the
distinct eigenvalues $\lambda_{P,L_y,j}$ and $\lambda_{Z,L_y,j}$ for
sufficiently large integer $q$, they may be zero or negative for some positive
integer values of $q$.  More generally, while they play the role of eigenvalue
multiplicities for sufficiently large integer $q$, the domain of their
definition may be generalized to $q \in {\mathbb R}_+$ or, indeed, to $q \in
{\mathbb C}$.

The dimension of the space of coloring configurations, ${\cal N}$, is equal to
the sum of the multiplicities of each distinct eigenvalue, i.e., the sum of the
dimensions of the invariant subspaces corresponding to each of these distinct
eigenvalues.  For the chromatic polynomial, this is ${\cal N}$, which is equal
to the sum 
\beq
C_{P,L_y}=\sum_{j=1}^{N_{P,L_y,\lambda}} c_{P,L_y,j}
\label{cpsum}
\eeq
while for the full Potts model partition function, we shall denote it as
\beq
C_{Z,L_y}=\sum_{j=1}^{N_{Z,L_y,\lambda}} c_{Z,L_y,j}  \ . 
\label{czsum}
\eeq
Some previous literature on chromatic polynomials and Potts model partition 
functions on lattice strips is in \cite{bds}-\cite{hd}. 

\section{Structural Theorems} 

\subsection{Coefficients}

We find that the coefficients that enter into (\ref{pgsum}) for $P(G_D,q)$ and
(\ref{zgsum}) for $Z(G_D,q,v)$ are polynomials in $q$ that consist of a special
restricted set with the property that the coefficient of maximal degree $d \ge
1$ in $q$ is
\beqs
\kappa^{(d)} & = & 2\Bigg [ U_{2d}\Big ( \frac{\sqrt{q}}{2} \Big ) -
T_{2d}\Big ( \frac{\sqrt{q}}{2} \Big ) \Bigg ] \cr\cr
& = & \sqrt{q} \ U_{2d-1} \Big ( \frac{\sqrt{q}}{2} \Big ) \cr\cr
& = & \sum_{j=0}^{d-1} (-1)^j { 2d-1-j \choose j} q^{d-j} 
\label{kappad}
\eeqs
where $T_n(x)$ and $U_n(x)$ are the Chebyshev polynomials of the first and
second kinds, defined by 
\beq
T_n(x) = \frac{1}{2}\sum_{j=0}^{[\frac{n}{2}]} (-1)^j \frac{n}{n-j}
{n-j \choose j} (2x)^{n-2j}
\label{tndef}
\eeq
and
\beq
U_n(x) = \sum_{j=0}^{[\frac{n}{2}]} (-1)^j {n-j \choose j} (2x)^{n-2j}
\label{undef}
\eeq
where in eqs. (\ref{tndef}) and (\ref{tndef}) 
$[\frac{n}{2}]$ in the upper limit on the summand means the integral part
of $\frac{n}{2}$. The first few of these coefficients are
\beq
\kappa^{(1)}=q
\label{kappa1}
\eeq
\beq
\kappa^{(2)}=q(q-2)
\label{kappa2}
\eeq
\beq
\kappa^{(3)}=q(q-1)(q-3)
\label{kappa3}
\eeq
\beq
\kappa^{(4)}=q(q-2)(q^2-4q+2)
\label{kappa4}
\eeq
\beq
\kappa^{(5)}=q(q^2-3q+1)(q^2-5q+5)
\label{kappa5}
\eeq
\beq
\kappa^{(6)}=q(q-1)(q-2)(q-3)(q^2-4q+1) \ . 
\label{kappa6}
\eeq
We have established the following factorization:
\beq
\kappa^{(d)} = \prod_{k=1}^d (q - s_{d,k})
\label{kappafactors}
\eeq
where
\beqs
s_{d,k} & = & 2 + 2\cos \bigg (\frac{\pi k}{d} \bigg ) \cr\cr
        & = & 4\cos^2 \bigg ( \frac{\pi k}{2d} \bigg ) \quad {\rm for} \ 
 k=1,2,..d
\label{skd}
\eeqs

In \cite{cf}, we showed that the coefficients $c_{L_y,j}$ that entered into
$Z(G_s,q,v)$ and hence also $P(G_s,q)$ for cyclic strips $G_s$of the square or
triangular lattices (and M\"obius strips of the square lattice) were of the
form \cite{cf}
\beqs
c^{(d)} & = & U_{2d}\Bigl ( \frac{\sqrt{q}}{2}\Bigr ) \cr\cr
        & = & \sum_{j=0}^d (-1)^{j}{2d-j \choose j}q^{d-j} \cr\cr
        & = & \prod_{k=1}^d (q - q_{d,k})
\label{cd}
\eeqs
where
\beq
q_{d,k} \equiv 2+2\cos \Bigl ( \frac{2\pi k}{2d+1} \Bigr )
 = 4\cos^2 \Bigl ( \frac{\pi k}{2d+1} \Bigr ) \quad {\rm for} \ 
k=1,2,...d \ .
\label{cdzeros}
\eeq
The $\kappa^{(d)}$ can be expressed as differences of these $c^{(d)}$
coefficients: 
\beq
\kappa^{(d)}=c^{(d)}-c^{(d-1)} \quad {\rm for} \ d = 1,2,...
\label{kappacd}
\eeq

The $\kappa^{(d)}$ coefficients have the following general properties.
\begin{itemize}

\item

Given the expression of $\kappa^{(d)}$, (\ref{kappad}) in terms of Chebyshev 
polynomials, and the recursion relation for these Chebyshev polynomials 
(for $n \ge 1$) 
\beq 
Ch_{n+1}(x)=2x Ch_n(x)- Ch_{n-1}(x) \quad {\rm for} \ Ch_n(x) = T_n(x) \
{\rm or} \ U_n(x)
\label{trecursion}
\eeq
we obtain the recursion relation for the $\kappa^{(d)}$ for $d \ge 2$, 
\beq
\kappa^{(d+1)} = (q-2)\kappa^{(d)}-\kappa^{(d-1)} \ . 
\label{kapparecursion}
\eeq

\item

$\kappa^{(d)}$ is a polynomial of degree $d$ in $q$ whose coefficients
alternate in sign.

\item

$\kappa^{(d)}$ has the factor $q$.

\item

If and only if $d$ is even and $d \ge 2$, then $\kappa^{(d)}$ has the factor
$(q-2)$.

\item

If and only if $d=0$ mod 3 and $d \ge 3$, then $\kappa^{(d)}$ has the factor
$(q-3)$.

\item

The coefficient of the term $q^d$ in $\kappa^{(d)}$ is 1.

\item

The coefficient of the term $q$ in $\kappa^{(d)}$ is $(-1)^{d+1}d$.

\item

The $d$ zeros of $\kappa^{(d)}$ occur at the (real) values $s_{d,k}$,
$k=1,...,d$, where $s_{d,k}$ was given in eq. (\ref{skd}).  Clearly, these lie
in the interval $0 \le q < 4$.  In passing, we note that the $k=2$ special 
case $s_{d,2}$ is the Tutte-Beraha number $B_d$.  

\end{itemize}

If one lets
\beq
q = 2 + 2 \cos \theta = 4\cos^2 \Bigl ( \frac{\theta}{2} \Bigr )
\label{qtrel}
\eeq
one sees that the argument of the Chebyshev polynomials in (\ref{kappad})
is given by $\frac{\sqrt{q}}{2}=\cos (\theta/2)$.  Using the 
identities (with $\omega=\theta/2$ here)
\beq
T_n(\cos \omega) = \cos(n \omega)
\label{tncos}
\eeq
and
\beq
U_n(\cos \omega) = \frac{\sin[(n+1)\omega]}{\sin \omega} 
\label{unsin}
\eeq
we have an alternate formula for $\kappa^{(d)}$, 
\beq
\kappa^{(d)} = 2\cot \omega \sin (2 d \omega)
\label{kappa_alt}
\eeq

\subsection{Determination of $n_P(L_{\lowercase{y}},d)$ and 
$N_{P,L_y,\lambda}$} 

Let us define $n_P(L_y,d)$ as the number of terms $\lambda_{P,L_y,j}$ in
$P(G_D(L_y \times L_x),q)$ that have as their coefficients
$c_{P,L_y,j}=\kappa^{(d)}$.  For $G_D(L_y \times L_x)$ strips, these
coefficients are nonzero for $1 \le d \le L_y+1$. (In our earlier work
\cite{cf} for different strip graphs, the coefficients were defined to start
with $c^{(0)}=1$, so that the range went from $0 \le d \le L_y$; in both cases,
the resultant chromatic polynomial has terms in $q$ ranging from $q^n$ to $q$.)
The total number, $N_{P,L_y,\lambda}$, of different terms $\lambda_{P,L_y,j}$
in (\ref{pgsum}) is then given by
\beq
N_{P,L_y,\lambda} = \sum_{d=1}^{L_y+1} n_P(L_y,d)  \ . 
\label{npsum}
\eeq
For the sum, (\ref{cpsum}), of the coefficients in (\ref{pgsum}) we have
\beq
C_{P,L_y} = \sum_{j=1}^{N_{P,L_y,\lambda}} c_{P,L_y,j} =
\sum_{d=1}^{L_y+1} n_P(L_y,d)\kappa^{(d)} \ . 
\label{cdsum}
\eeq

Using the same methods as in \cite{cf}, we have determined the numbers
$n_P(L_y,d)$ of $\lambda_{P,L_y,j}$'s that have each type of coefficient,
$\kappa^{(d)}$.  From coloring matrix arguments, using the fact that a
transverse slice along the strip is a path (tree) graph $T_{L_y+1}$, for which
the chromatic polynomial is $P(T_{L_y+1},q)=q(q-1)^{L_y}$, it follows that
\beq
\sum_{d=1}^{L_y+1} n_P(L_y,d)\kappa^{(d)} = P(T_{L_y+1},q)=q(q-1)^{L_y} \ . 
\label{npsum_dbc2}
\eeq
Differentiating eq. (\ref{npsum_dbc2}) $L_y$ times we obtain $L_y+1$ linear
equations in the $L_y+1$ unknowns $n_P(L_y,d)$ for $1 \le d \le L_y+1$.
Solving these equations, we find that 
\beq
n_P(L_y,d)=0 \quad {\rm for} \quad d > L_y+1
\label{npup}
\eeq
\beq
n_P(L_y,L_y+1)=1
\label{npcly}
\eeq
and
\beq
n_P(1,1)=1
\label{np11}
\eeq
with all other numbers $n_P(L_y,d)$ being determined by the recursion
relation
\beq
n_P(L_y+1,d) = n_P(L_y,d-1)+n_P(L_y,d)+n_P(L_y,d+1)
\quad {\rm for} \quad 1 \le d \le L_y+1 \ . 
\label{nprecursion}
\eeq
In particular, we find 
\beq
n_P(L_y,L_y)=L_y
\label{npclyminus1}
\eeq
and
\beq
n_P(L_y,1)=M_{L_y}
\label{nply1}
\eeq
where $M_n$ is the Motzkin number, given by 
\beq
M_n =  \sum_{j=0}^n (-1)^j C_{n+1-j} {n \choose j}
\label{motzkin}
\eeq
and 
\beq
C_n=\frac{1}{n+1}{2n \choose n}
\label{catalan}
\eeq
is the Catalan number. (The Catalan and Motzkin numbers occur in many 
combinatoric applications \cite{motzkin}-\cite{stanley}.) 

We find that the $n_P(L_y,d)$ are closely related to certain numbers
enumerating random walks.  To explain this, consider a random walk on the
nonnegative integers such that in each step the walker moves by $+1$, $-1$, or
0 units.  Denote $m(n,k)$ as the number of walks of length $n$ steps starting
at $0$ and ending at $k$.  The case $k=0$ describes the number of
walks defined above that return to the origin after $n$ steps. This is given by
the Motzkin number; $m(n,0)=M_n$. Define the sum
\beq
{\cal S}_n = \sum_{k=0}^n  m(n,k) \ . 
\label{rowsum}
\eeq
We find that the numbers $n_P(L_y,d)$ are
precisely equal to these enumerations of random walks: 
\beq
n_P(L_y,d)=m(L_y,d-1) \quad {\rm for} \ 1 \le d \le L_y+1
\label{motzkintri}
\eeq
The values of the $n_P(L_y,d)$ for $L_y=1$ to $L_y=10$ are listed in Table
\ref{nptable}. (Note that ${\cal S}_9$ should read 6046 in the related Table 3 
of Ref. \cite{cf}.) 

 From the identity (\ref{motzkintri}), it follows that the total number of
terms $N_{P,L_y,\lambda}$ is 
\beq
N_{P,L_y,\lambda} = {\cal S}_{L_y} \ . 
\label{nptots}
\eeq
These numbers are listed in Table \ref{nptable}.

\begin{table}
\caption{\footnotesize{Table of numbers $n_P(L_y,d)$ and their sums,
$N_{P,L_y,\lambda}$ for $G_D(L_y \times L_x)$. Blank entries are zero.}}
\begin{center}
\begin{tabular}{|c|c|c|c|c|c|c|c|c|c|c|c|c|} \hline\hline
$L_y \ \downarrow$ \ \ $d \ \rightarrow$
  & 1 & 2   & 3   & 4   & 5   & 6  & 7  & 8 & 9 & 10 & 11 & 
$N_{P,L_y,\lambda}$ \\ \hline\hline
1  & 1   & 1   &     &     &     &     &     &    &   &    &   & 2    \\ \hline
2  & 2   & 2   & 1   &     &     &     &     &    &   &    &   & 5    \\ \hline
3  & 4   & 5   & 3   & 1   &     &     &     &    &   &    &   & 13   \\ \hline
4  & 9   & 12  & 9   & 4   & 1   &     &     &    &   &    &   & 35   \\ \hline
5  & 21  & 30  & 25  & 14  & 5   & 1   &     &    &   &    &   & 96   \\ \hline
6  & 51  & 76  & 69  & 44  & 20  & 6   & 1   &    &   &    &   & 267  \\ \hline
7  & 127 & 196 & 189 & 133 & 70  & 27  & 7   & 1  &   &    &   & 750  \\ \hline
8  & 323 & 512 & 518 & 392 & 230 & 104 & 35  & 8  & 1 &    &   & 2123 \\ \hline
9  & 835 & 1353& 1422& 1140& 726 & 369 & 147 & 44 & 9 & 1  &   & 6046 \\ \hline
10 & 2188& 3610& 3915& 3288& 2235& 1242& 560 & 200& 54& 10 & 1 & 17303 \\ 
\hline\hline
\end{tabular}
\end{center}
\label{nptable}
\end{table}

We have determined a generating function whose expansion yields the numbers 
$N_{P,L_y,\lambda}$:
\beq
\frac{1}{2}\biggl [ \biggl ( \frac{1+x}{1-3x} \biggr )^{1/2} - 1 \biggr ] - x =
\sum_{L_y=1}^\infty N_{P,L_y,\lambda}x^{L_y+1} \ . 
\label{genfun}
\eeq
Combining this with our results from \cite{cf}, we observe the interesting fact
that the total numbers of terms $N_{P,L_y,\lambda}$ for the chromatic
polynomial of a $G_D(L_y \times L_x)$ strip graph (for any $L_x$) is equal to
1/2 the corresponding total number of terms for the cyclic or M\"obius strips
of the square or triangular lattice with the next larger width, $L_y+1$, i.e.,
\beq
N_{P,L_y,\lambda} = \frac{1}{2}N_{P,\Lambda,cyc/Mb,L_y+1} \quad \Lambda=sq,tri 
\ . 
\label{nptotrel}
\eeq
Further, we observe the intriguing relation 
\beq
N_{P,L_y,\lambda} = N_{DA,sq,L_y+1}
\label{animals_p}
\eeq
where $N_{DA,sq,n}$ denotes the number of directed $n$-site lattice 
animals on the square lattice.  We prove this by noting that the generating
function (\ref{genfun}) that we have found for $N_{P,L_y,\lambda}$ is the same
as the generating function for directed $n$-site lattice animals on the square
lattice given in \cite{animal}.

  From (\ref{genfun}), it follows that for large width $L_y$,
$N_{P,L_y,\lambda}$ grows exponentially fast, with the leading asymptotic
behavior:
\beq
N_{P,L_y,\lambda} \sim L_y^{-1/2} \ 3^{L_y} \quad {\rm as} \ \ L_y \to \infty \
. 
\label{npasymp}
\eeq
This is the same asymptotic behavior that we found in \cite{cf} for the total
number of terms entering in the chromatic polynomials for the cyclic or
M\"obius strips of the square or triangular lattice of width $L_y$. 

\subsection{Determination of $n_Z(L_{\lowercase{y}},d)$ and 
$N_{Z,L_y,\lambda}$} 

 From the same type of coloring matrix arguments used in \cite{cf}, we have 
\beq
\sum_{d=1}^{L_y+1} n_Z(L_y,d)\kappa^{(d)} = q^{L_y+1} \ . 
\label{nzsum_dbc2}
\eeq
Differentiating eq. (\ref{nzsum_dbc2}) $L_y$ times we obtain $L_y+1$ linear
equations in the $L_y+1$ unknowns $n_Z(L_y,d)$ for $1 \le d \le L_y+1$.
Solving these equations, we obtain the general results
\beq
n_Z(L_y,d)=0 \quad {\rm for} \quad d > L_y+1
\label{nzup}
\eeq
\beq
n_Z(L_y,L_y+1)=1
\label{nzcly}
\eeq
and
\beq
n_P(1,1)=2
\label{nz11}
\eeq
with all other numbers $n_Z(L_y,d)$ being determined by the recursion
relation
\beq
n_Z(L_y+1,d) = n_Z(L_y,d-1)+2n_Z(L_y,d)+n_Z(L_y,d+1)
\quad {\rm for} \quad 1 \le d \le L_y+1 \ . 
\label{nzrecursion}
\eeq
We solve this recursion relation with the conditions
(\ref{nzup})-(\ref{nz11}) in closed form and obtain 
\beq
n_Z(L_y,d)=\frac{2d}{L_y+d+1}{2L_y+1 \choose L_y-d+1} \ . 
\label{nzLydGd}
\eeq
Summing these numbers, we find, for the total number of terms 
\beq
N_{Z,L_y,\lambda} = { 2L_y+1 \choose L_y+1} \ . 
\label{nztot}
\eeq

We observe that the total numbers of terms $N_{Z,L_y,\lambda}$ for the Potts
model partition function on the $G_D(L_y \times L_x)$ strip graph 
(for any $L_x$) is equal to 1/2 the corresponding total number of terms for 
the cyclic or M\"obius strips of the square or triangular lattice with the 
next larger width, $L_y+1$, i.e.,
\beq
N_{Z,L_y,\lambda} = \frac{1}{2}N_{Z,\Lambda,cyc/Mb,L_y+1} \quad \Lambda=sq,tri
\ . 
\label{nztotrel}
\eeq
There is also an interesting connection with directed lattice animals: 
\beq
N_{Z,L_y,\lambda} = N_{DA,tri,L_y+1}
\label{animals_z}
\eeq
where $N_{DA,tri,n}$ denotes the number of directed $n$-site lattice 
animals on the triangular lattice. 

As $L_y \to \infty$, $N_{Z,L_y,\lambda}$ has the leading asymptotic behavior
\beq
N_{Z,L_y,\lambda} \sim \pi^{-1/2}L_y^{-1/2} \ 4^{L_y} \quad
{\rm as} \ \ L_y \to \infty \ . 
\label{nztotasymp}
\eeq
The values of $n_Z(L_y,d)$ and $N_{Z,L_y,d}$ are given for $1 \le L_y \le 10$
in Table \ref{nztable}. 

\begin{table}
\caption{\footnotesize{Table of numbers $n_Z(L_y,d)$ and their sums,
$N_{Z,L_y,\lambda}$ for $G_D(L_y \times L_x)$. Blank entries are zero.}}
\begin{center}
\begin{tabular}{|c|c|c|c|c|c|c|c|c|c|c|c|c|}
\hline\hline
$L_y \ \downarrow$ \ \ $d \ \rightarrow$
  & 1 & 2   & 3   & 4   & 5   & 6  & 7  & 8 & 9 & 10 & 11 & $N_{P,L_y,\lambda}$
\\ \hline\hline
1  & 2   & 1   &     &     &     &     &     &    &   &    &   & 3   \\ \hline
2  & 5   & 4   & 1   &     &     &     &     &    &   &    &   & 10  \\ \hline
3  & 14  & 14  & 6   & 1   &     &     &     &    &   &    &   & 35  \\ \hline
4  & 42  & 48  & 27  & 8   & 1   &     &     &    &   &    &   & 126  \\ \hline
5  & 132 & 165 & 110 & 44  & 10  & 1   &     &    &   &    &   & 462  \\ \hline
6  & 429 & 572 & 429 & 208 & 65  & 12  & 1   &    &   &    &   & 1716 \\ \hline
7  & 1430& 2002& 1638& 910 & 350 & 90  & 14  & 1  &   &    &   & 6435 \\ \hline
8  & 4862& 7072& 6188& 3808& 1700& 544 & 119 & 16 & 1 &    &   & 24310\\ \hline
9  &16796&25194&23256&15504& 7752& 2907& 798 & 152& 18& 1  &   & 92378\\ \hline
10 &58786&90440&87210&62016&33915&14364& 4655&1120&189& 20 & 1 & 352716 \\
\hline\hline
\end{tabular}
\end{center}
\label{nztable}
\end{table}

In what follows, we shall need to refer to our calculation of the numbers
$n_Z(L_y,d)$ of $\lambda$'s with associated coefficient $c^{(d)}$ for cyclic
strips $G_s$ of the square or triangular lattice of width $L_y$ (and
arbitrarily $L_x$ in \cite{cf}.  We found
\beq
n_Z(G_s,L_y,d)=\frac{2d+1}{L_y+d+1}{2L_y \choose L_y-d}
\label{nzLydcyc}
\eeq
for $0 \le d \le L_y$ and zero otherwise.  These values will be needed below
and are listed in Table \ref{nztablecyc}. 

\begin{table}
\caption{\footnotesize{Table of numbers $n_Z(L_y,d)$ and their sums,
$N_{Z,G,\lambda}$ for cyclic strips $G_s(L_y \times L_x)$ of the square or
triangular lattice of width $L_y$ and arbitrary length $L_x$
\protect{\cite{cf}}.}}
\begin{center}
\begin{tabular}{|c|c|c|c|c|c|c|c|c|c|c|c|c|}
\hline\hline
$L_y \ \downarrow$ \ \ $d \ \rightarrow$
   & 0 & 1   & 2   & 3   & 4   & 5  & 6  & 7 & 8 & 9 & 10 &
$N_{Z,L_y,\lambda}$
\\ \hline\hline
1  & 1   & 1   &     &     &     &    &    &   &   &   &   & 2     \\ \hline
2  & 2   & 3   & 1   &     &     &    &    &   &   &   &   & 6     \\ \hline
3  & 5   & 9   & 5   & 1   &     &    &    &   &   &   &   & 20    \\ \hline
4  & 14  & 28  & 20  & 7   & 1   &    &    &   &   &   &   & 70    \\ \hline
5  & 42  & 90  & 75  & 35  & 9   & 1  &    &   &   &   &   & 252   \\ \hline
6  & 132 & 297 & 275 & 154 & 54  & 11 & 1  &   &   &   &   & 924   \\ \hline
7  & 429 & 1001& 1001& 637 & 273 & 77 & 13 & 1 &   &   &   & 3432  \\ \hline
8  & 1430& 3432& 3640& 2548& 1260& 440& 104& 15& 1 &   &   & 12870 \\ \hline
9  & 4862&11934&13260& 9996& 5508&2244& 663&135& 17& 1 &   & 48620 \\ \hline
10 &16796&41990&48450&38760&23256&10659&3705&950&170&19& 1 & 184756 \\
\hline\hline
\end{tabular}
\end{center}
\label{nztablecyc}
\end{table}

\section{Connection with Bratteli Diagrams for $TL_{\lowercase{n}}
({\lowercase{q}})$}

There are a number of interesting algebraic properties of structural elements
and properties of the chromatic polynomials and Potts model partition functions
equivalently Tutte polynomials.  The interpretation of the coefficients
$c_{P,G_s,j}$ in chromatic polynomials as dimensions of invariant subspaces of
coloring matrices was discussed in \cite{b,matmeth} and its generalization to
the full Potts model partition function was discussed in \cite{cf}.  Thus,
$C_{P,G_s,L_y}$ and $C_{Z,G_s,L_y}$ are sums of these dimensions, a fact that
we have used in \cite{cf} and eqs. (\ref{npsum_dbc2}) and (\ref{nzsum_dbc2})
here.

In \cite{cf} we pointed out another connection, namely the fact that the
numbers $n_Z(G_s,L_y,d)$ for $0 \le d \le L_y$ for cyclic strips $G_s$ of the
square or triangular lattice of width $L_y$ were related to Temperley-Lieb
algebras.  We have found a very interesting generalization of this for the
$n_Z(G_D, L_y,d)$ calculated here.  Let us denote $TL_n(q)$ as the
Temperley-Lieb algebra of operators $U_i$, $n=1,..,n$ satisfying
\cite{tl,jones89}
\beq
U_i^2 = q^{1/2} U_i
\label{tl1}
\eeq
\beq
[U_i,U_j]=0 \quad {\rm if} \ |i-j| \ne 1
\label{tl2}
\eeq
where $[A,B]=AB-BA$, and 
\beq
U_i U_{i\pm1} U_i = U_i
\label{tl3}
\eeq
Associated with the structural decomposition of this algebra $TL_n(q)$ there is
what is known as a Bratteli diagram \cite{bratteli,jones89,ghj}, given in Table
\ref{bratteli}.

\begin{table}
\caption{\footnotesize{Generic Bratteli diagram relevant for $TL_n(q)$.}}
\begin{center}
\begin{tabular}{|c|c|c|c|c|c|c|c|c|c|c|c|}
\hline\hline
$n \ \downarrow$ \ \ $m \ \rightarrow$
  & 1 & 2   & 3   & 4   & 5   & 6  & 7  & 8 & 9 & 10 & 11 \\ \hline\hline
1  &     & 1   &     &     &     &     &     &    &   &    &    \\ \hline
2  &  1  &     & 1   &     &     &     &     &    &   &    &    \\ \hline
3  &     & 2   &     & 1   &     &     &     &    &   &    &    \\ \hline
4  &  2  &     & 3   &     & 1   &     &     &    &   &    &    \\ \hline
5  &     & 5   &     & 4   &     & 1   &     &    &   &    &    \\ \hline
6  &  5  &     & 9   &     & 5   &     & 1   &    &   &    &    \\ \hline
7  &     & 14  &     & 14  &     & 6   &     & 1  &   &    &    \\ \hline
8  & 14  &     & 28  &     & 20  &     & 7   &    & 1 &    &    \\ \hline
9  &     & 42  &     & 48  &     & 27  &     & 8  &   & 1  &    \\ \hline
10 & 42  &     & 90  &     & 75  &     & 35  &    & 9 &    & 1  \\
\hline\hline
\end{tabular}
\end{center}
\label{bratteli}
\end{table}
We observe the following interesting properties:

\begin{itemize}

\item 

For even $n=2L_y$, the entries in this row of the Bratteli diagram
are equal to the $n_Z(L_y,d)$ for the cyclic strip of the square lattice of
width $L_y$ and arbitrary length $L_x$ for $m=2d+1$ with $d$ in the range 
$0 \le d \le L_y$, i.e., $1 \le m \le 2L_y+1 = n+1$. 

\item 

For odd $n=2L_y+1$, the entries in this row  of the Bratteli diagram
are equal to the $n_Z(L_y,d)$ for the strip $G_D$ of width $L_y$ and arbitrary
length $L_x$ for $m=2d$ with  $d$ in the range $1 \le d \le L_y+1$, i.e., 
$2 \le m \le 2(L_y+1) = n+1$. 

\end{itemize}

As this makes clear, the numbers $n_Z(L_y,d)$ for the cyclic strip and the
$G_D$ strip can be expressed via a single formula.  Define
\beq
C_{n,m}={n-1\choose m} - {n-1 \choose m-2}
\label{cnp}
\eeq
so that the $C_{n,m}$ satisfy the recursion relation
\beq
C_{n,m}=C_{n-1,m-1} + C_{n-1,m+1} \ . 
\label{cnprecursion}
\eeq
For $G_{cyc}(L_y \times L_x)$, let $n=2L_y$ and $m=L_y-d$ with $0 \le d \le
L_y$.  For $G_D(L_y \times L_x)$, let $n=2L_y+1$ and $m=L_y+1-d$ with $1 \le
d \le L_y+1$.  Then with these substitutions, $C_{n,m}$ reduces to the
respective expressions (\ref{nzLydcyc}) and (\ref{nzLydGd}).

Indeed, the fact that our results in \cite{cf} related $n_Z(L_y,d)$ for the
cyclic strips of the square and triangular lattices to the even-$n$ rows of the
Bratteli diagram, i.e., the rows for $TL_{2L_y}(q)$, raised a question: is
there a family of graphs with the property that the coefficients in the generic
formula (\ref{zgsum}) are restricted to a simple set such that the number of
$\lambda$'s with coefficients of a particular type fill out the odd-$n$ rows of
the Bratteli diagram.  We had searched for such a family of graphs and had
finally found it in the form of the $G_D(L_y \times L_x)$ graphs, which have
provided an affirmative answer to this question: the entries in the odd
$n=2L_y+1$ rows are given by the $n_Z(L_y,d)$, as described above.  Thus, our
current results are of interest both for the connection with Bratteli diagrams
and for the fact that, together with our previous findings in \cite{cf}, they
saturate this connection, i.e., the entries in all of the Bratteli diagram are
now accounted for in terms of the 1-1 correspondence with $n_Z(L_y,d)$ numbers
for these cyclic and $G_D$ strip graphs.

We observe that the Bratteli diagram also has a combinatoric interpretation.
Consider random walks on the non-negative integers that begin at the origin and
have the property that the walker moves by $+1$ or $-1$ spaces at each
step. The $(n,m)$ entry in the Bratteli diagram is the number of random walks
of this type, consisting of $n$ steps, that end at $m-1$. Note that the number
of random walks of this type that consist of $2n$ steps (a misprint in
\cite{cf} had this as $n$ steps) is the Catalan number $C_n$. The recursion
relation (\ref{cnprecursion}) follows immediately from this.

\section{Width $L_{\lowercase{y}}=1$}

It is of interest to calculate Potts model partition functions and chromatic
polynomials for these families of self-dual graphs.  In this section we shall
discuss the chromatic polynomial for the lowest case, $L_y=1$, and in the next
sections we shall present calculations of the chromatic polynomials for $2 \le
L_y \le 4$ and shall analyze the zeros of these polynomials in the complex $q$
plane and their accumulation sets as $L_x \to \infty$.  In a separate
publication we shall present more lengthy calculations of the full Potts model
partition functions. 

We recall that the $L_y=1$ family is comprised of the wheel graphs $(Wh)$ with
$L_x$ vertices on the rim and a central vertex, with edges forming ``spokes''
connecting this central graph to the vertices on the rim.  The $n$-vertex wheel
graph can be constructed as the ``join'' $(Wh)_{n+1}= K_1 + C_n$, where $K_n$
and $C_n$ are the complete graph and the circuit graph on $n$ vertices,
respectively.  (The join of two graphs $G$ and $H$ is defined as the graph
obtained by connecting each vertex of $G$ to every vertex of $H$ by edges. The
complete graph $K_n$ is defined as the graph with $n$ vertices such that each
vertex is adjacent to every other vertex.)  The chromatic polynomial has the
form of (\ref{pgsum}) with $N_{P,L_y=1,\lambda}=2$ and the terms
\beq
\lambda_{1,1}=-1
\label{lam1s1}
\eeq
\beq
\lambda_{1,2}=q-2
\label{lam2s1}
\eeq
and coefficients
\beq
c_{1,1}=\kappa^{(2)}
\label{c1s1}
\eeq
\beq
c_{1,2}=\kappa^{(1)} \ . 
\label{c2s1}
\eeq
Here and below for brevity we write $\lambda_{L_y,j}$ for $\lambda_{P,L_y,j}$
and $c_{L_y,j}$ for $c_{P,L_y,j}$.  Thus, 
\beq
P(1 \times L_x, q) = (Wh)_{L_x+1} = q\bigg [ (q-2)(-1)^{L_x} + (q-2)^{L_x} 
\bigg ] \ . 
\label{pwheel}
\eeq
The continuous locus ${\cal B}$ formed as the continuous accumulation set of
chromatic zeros in the limit $L_x \to \infty$ limit, is the unit circle
centered at $q=2$ \cite{w,wc}
\beq
{\cal B}: \ |q-2|=1 \quad {\rm for} \ L_y=1 \ . 
\label{bwheel}
\eeq
Evidently, this intersects the real axis at $q=1$ and $q=3$.  The locus ${\cal
B}$ separates the $q$ plane into two regions, $R_1$ and $R_2$, which are,
respectively, the exterior and the interior of the circle (\ref{bwheel}). 
In region $R_1$ 
\beq
W = q-2 \quad {\rm for} \ q \in R_1 \ . 
\label{wwheel_r1}
\eeq
In any other region than $R_1$, only the magnitude $|W|$ can be determined
unambiguously \cite{w}, and we have
\beq
|W|=1 \quad {\rm for} \ q \in R_2 \ . 
\label{wwheel_r2}
\eeq
Except for the single discrete zero at $q=2$, the chromatic zeros lie exactly
on the circle $|q-2|=1$ and occur with a density that is a constant as a
function of angular position on this circle \cite{wc}. 

\section{Width $L_{\lowercase{y}}=2$}

Our general formula (\ref{genfun}) yields the result that there are 
$N_{P,L_y=2,\lambda}=5$ terms for this width $L_y=2$.  The chromatic
polynomial has the form (\ref{pgsum}) with the following terms, listed in order
of descending degree of their associated coefficients. 
\beq
\lambda_{2,1} = 1
\label{lam1s2}
\eeq
\beq
\lambda_{2,(2,3)} = \frac{1}{2}(5-2q \pm \sqrt{5} \ )
\label{lam23s2}
\eeq
\beq
\lambda_{2,(4,5)} = \frac{1}{2}\bigg [q^2-5q+7 \pm (q^4-6q^3+15q^2-22q+17
)^{1/2} \bigg ]
\label{lam45s2}
\eeq
and the corresponding coefficients
\beq
c_{2,1} = \kappa^{(3)}
\label{c1s2}
\eeq
\beq
c_{2,j}=\kappa^{(2)} \quad {\rm for} \ j=2,3
\label{c23s2}
\eeq
\beq
c_{2,j}=\kappa^{(1)}  \quad {\rm for} \ j=4,5 \ . 
\label{c45s2}
\eeq

\begin{figure}[hbtp]
\centering
\leavevmode
\epsfxsize=4.0in
\begin{center}
\leavevmode
\epsffile{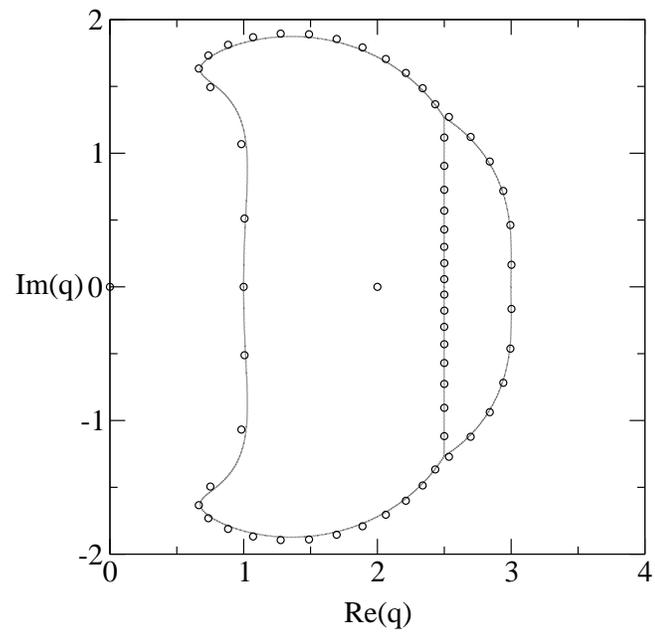}
\end{center}
\caption{\footnotesize{Singular locus ${\cal B}$ for the $L_x \to \infty$ limit
of $G_D( 2 \times L_x)$. For comparison, chromatic zeros are shown for 
$L_x=30$ (i.e., $n=61$).}}
\label{whpxy2}
\end{figure}

The locus ${\cal B}$ is shown in Fig. \ref{whpxy2}, together with chromatic
zeros for a long finite strip.  Evidently, the chromatic zeros (except the
discrete zero at $q=2$), which eventually merge to form this locus in the $L_x
\to \infty$ limit, generally lie close to the curves on ${\cal B}$.  This locus
divides the $q$ plane into three regions: (i) region $R_1$ including the
intervals $q > 3$ and $q < 1$ on the real axis and extending outward to
infinitely large $|q|$; (ii) the region $R_2$ including the real interval $5/2
< q < 3$, and (iii) the innermost region $R_3$ including the real interval $1 <
q < 5/2$.  There are triple points at which all three regions are contiguous at
$q = q_t,q_t^*$, where $q_t \simeq 5/2 + 1.27 i$.  Note that the part of the
boundary ${\cal B}$ separating regions $R_2$ and $R_3$ is the line segment
$Re(q)=5/2$, $-Im(q_t) < Im(q) < Im(q_t)$. In region $R_1$, $\lambda_{2,4}$ is
dominant, so
\beq
W=(\lambda_{2,4})^{1/2} \quad {\rm for} \ q \in R_1
\label{ws2r1}
\eeq
\beq
|W| = |\lambda_{2,3}|^{1/2} \quad {\rm for} \ q \in R_2
\label{ws2r2}
\eeq
\beq
|W|=|\lambda_{2,2}|^{1/2} \quad {\rm for} \ q \in R_3 \ . 
\label{ws2r3}
\eeq
As is evident in Fig. \ref{whpxy2}, the locus ${\cal B}$ crosses the real $q$
axis at $q=1$, $q=5/2$, and $q=3$.

\section{Width $L_{\lowercase{y}}=3$}

Here our general formula (\ref{genfun}) yields $N_{P,L_y=3,\lambda}=13$.  We 
have calculated the chromatic polynomial and obtain the following results,
where again the $\lambda$'s are listed in order of descending degree of their
coefficients.  The first term is 
\beq
\lambda_{3,1} = -1 \ . 
\label{lam1s3}
\eeq
Next, the $\lambda_{3,j}$ for $2 \le j \le 4$ are the roots of the equation
\beq
\xi^3 - (3q-8)\xi^2 + (3q^2-16q+19)\xi - (q^3-8q^2+19q-13) = 0 \ . 
\label{eqLy3_cub}
\eeq
The $\lambda_{3,j}$ for $5 \le j \le 9$ are the roots of the equation 
\beqs
& & \xi^5 + (3q^2-16q+23)\xi^4 + (3q^4-34q^3+142q^2-261q+177)\xi^3 \cr\cr
& + & (q^6-20q^5+153q^4-591q^3+1234q^2-1332q+581)\xi^2 \cr\cr
& - & (2q^7-35q^6+257q^5-1028q^4+2423q^3-3371q^2+2566q-823)\xi \cr\cr
& + & (q-2)^2(q^6-14q^5+79q^4-230q^3+366q^2-304q+103) = 0 \ . 
\label{eqLy3_fifth}
\eeqs
Finally, the $\lambda_{3,j}$ for $10 \le j \le 13$ are the roots of the
equation 
\beqs
& & \xi^4 - (q^3-8q^2+24q-26)\xi^3 -(2q^5-23q^4+111q^3-279q^2+362q-191)\xi^2
\cr\cr
& - & (q-3)(q^6-14q^5+81q^4-251q^3+441q^2-415q+161)\xi \cr\cr
& + & (q-2)^3(q^5-11q^4+48q^3-104q^2+112q-47) = 0 \ . 
\label{eqLy3_fourth}
\eeqs
The corresponding coefficients are
\beq
c_{3,1}=\kappa^{(4)}
\label{c1s3}
\eeq
\beq
c_{3,j}=\kappa^{(3)} \quad {\rm for} \ 2 \le j \le 4
\label{c234s3}
\eeq
\beq
c_{3,j}=\kappa^{(2)} \quad {\rm for} \ 5 \le j \le 9
\label{c59s3}
\eeq
\beq
c_{3,j}=\kappa^{(1)} \quad {\rm for} \ 10 \le j \le 13 \ . 
\label{c1013s3}
\eeq

\begin{figure}[hbtp]

\centering
\leavevmode
\epsfxsize=4.0in
\begin{center}
\leavevmode
\epsffile{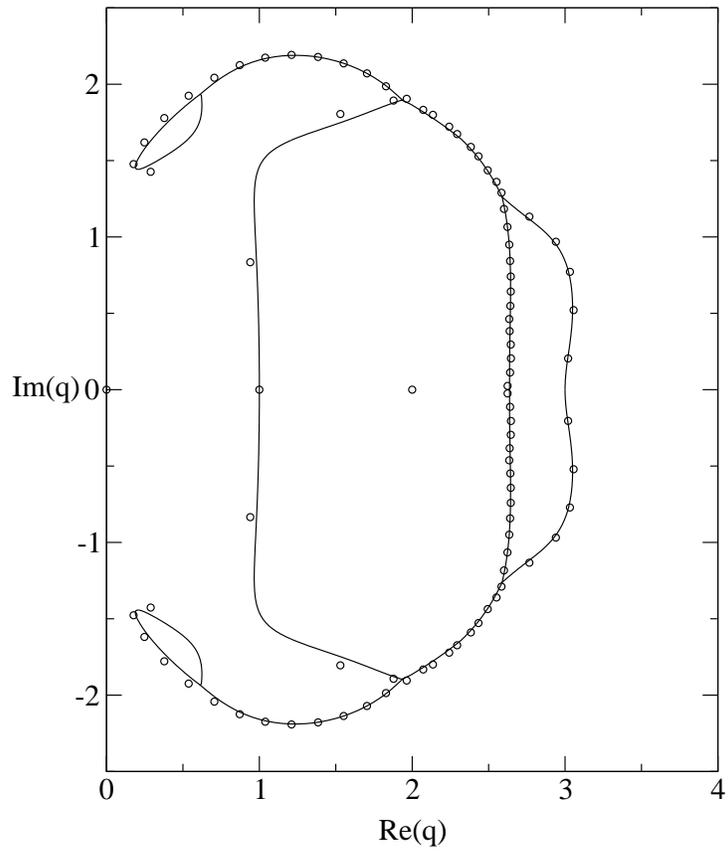}
\end{center}
\caption{\footnotesize{Singular locus ${\cal B}$ for the $L_x \to \infty$ limit
of $G_D(3 \times L_x)$. For comparison, chromatic zeros are shown for
$L_x=30$ (i.e., $n=91$).}}
\label{whpxy3}
\end{figure}

The locus ${\cal B}$ defined in the $L_x \to \infty$ limit is shown in
Fig. \ref{whpxy3}, together with chromatic zeros for a long finite strip.  This
locus divides the $q$ plane several regions, including (i) $R_1$, which
contains the real intervals $q < 1$ and $q > 3$ and extends outwards to
infinite $|q|$; (ii) $R_2$, which contains the real interval from $q \simeq
2.637$ to $q=3$; (iii) $R_3$, which contains the real interval $1 < q < 2.637$;
and (iv) the complex-conjugate pair of small regions $R_4$, $R_4^*$, centered
approximately at $q \simeq 0.44 \pm 1.67i$.  The density of chromatic zeros
varies in different parts of ${\cal B}$; this density is low on the portion of
${\cal B}$ that passes through $q=1$ and on the inner sides of the small
regions $R_4,R_4^*$.  In $R_1$, the dominant term is a root of the quartic
equation (\ref{eqLy3_fourth}), which we denote $\lambda_{3,R1}$, so that
\beq
W = (\lambda_{3,R1})^{1/3} \quad {\rm for} \ q \in R_1 \ . 
\label{ws3r1}
\eeq
In regions $R_2$ and $R_3$, the dominant terms are two different roots of the
fifth-degree equation (\ref{eqLy3_fifth}), and a root of this equation is also
dominant in the complex-conjugate pairs of regions $R_4$, $R_4^*$.  There are
three complex-conjugate pairs of triple points on ${\cal B}$.  Our experience
with the $L_x \to \infty$ limits of other families of strip graphs showed that
there can often be additional extremely small regions \cite{wcyl}; we have not
made an exhaustive search for these.

\section{$L_{\lowercase{y}}=4$} 

\begin{figure}[hbtp]
\centering
\leavevmode
\epsfxsize=4.0in
\begin{center}
\leavevmode
\epsffile{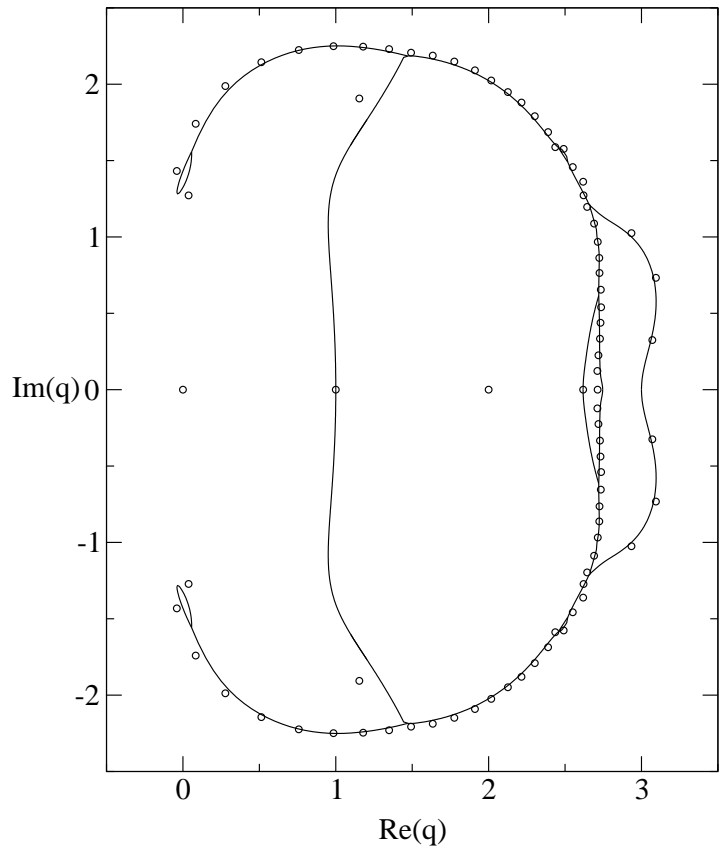}
\end{center}
\caption{\footnotesize{Singular locus ${\cal B}$ for the $L_x \to \infty$ limit
of $G_D(4 \times L_x)$. For comparison, chromatic zeros are shown for
$L_x=20$ (i.e., $n=81$).}}
\label{whpxy4}
\end{figure}

Here our general formula (\ref{genfun}) gives $N_{P,4,\lambda}=35$.  We
have 
\beq
\lambda_{4,1}=1
\label{lam1s4}
\eeq
with coefficient 
\beq
c_{4,1}=\kappa^{(5)}
\label{c1s4}
\eeq
and
\beq
\lambda_{4,2}=2-q \ . 
\label{lam2s4}
\eeq
The $\lambda_{4,j}$ for $3 \le j \le 5$ are roots of the equation
\beq
\xi^3+3(q-3)\xi^2-3(q-2)(q-4)\xi+(q^3-9q^2+24q-17)=0 \ . 
\label{lam4jcubic}
\eeq
These $\lambda_{4,j}$, $2 \le j \le 5$ have the coefficient
\beq
c_{4,j}=\kappa^{(4)} \quad {\rm for} \ 2 \le j \le 5 \ . 
\label{c25s4}
\eeq
The other 30 terms consist of roots of two equations of degree 9 and an
equation of degree 12.  Since these are rather lengthy, they are relegated to
the appendix of the copy of this paper in the cond-mat archive.  As before for
the other $G_D$ strips studied in this paper, these equations are in 1-1
correspondence with the coefficients listed in Table \ref{nptable}, so that all
of the roots of a given equation have the same coefficient, as specified in
this table.

The locus ${\cal B}$ defined in the $L_x \to \infty$ limit is shown in
Fig. \ref{whpxy4}, together with chromatic zeros for a long finite strip.  This
locus ${\cal B}$ divides the $q$ plane into a number of regions. These include
(i) $R_1$, which contains the real intervals $q < 1$ and $q > 3$ and extends
outwards to infinite $|q|$ as before; (ii) $R_2$, which contains the real
interval from $2.746 \lsim q < 3$; (iii) a narrow region $R_3$, which contains
the real interval $(1/2)(3+\sqrt{5}) \lsim q \lsim 2.746$, and (iv) $R_4$,
which contains the real interval $1 < q \lsim (1/2)(3+\sqrt{5})$; (v) the
complex-conjugate pair of small regions $R_5$, $R_5^*$, centered approximately
at $q \simeq \pm 1.4i$; and (vi) a complex-conjugate pair of tiny sliver
regions $R_6$, $R_6^*$ located around $q \simeq 5/2 \pm 1.55i$.  We note in
passing that $(1/2)(3+\sqrt{5})=B_5 = 2.61803...$, where $B_r=4\cos^2(\pi/r)$
is the Tutte-Beraha number.  The same caveat given above again applies; we have
not made an exhaustive search for other even tinier sliver regions.  The small
regions $R_5$ and $R_5^*$ extend into the $Re(q) < 0$ half-plane. In region
$R_1$ the dominant term is a root of the ninth-order equation with coefficient
$\kappa^{(1)}$ which we denote $\lambda_{4,R1}$, so that $W =
(\lambda_{4,R1})^{1/4}$.  The density of chromatic zeros is lowest on the
portions of ${\cal B}$ that pass through $q=1$ and $q = B_5$ and is
intermediate on the portion that passes through $q=3$.

\section{General Features} 

Our exact results exhibit a number of features: 

\begin{itemize}

\item 

For a given width $L_y$, the $\lambda_{Ly,j}$ with coefficient
$c_{L_y,j}=\kappa^{(d)}$, where $1 \le d \le L_y+1$, of which there are
$n_P(L_y,d)$, are roots of an algebraic equation of degree at most
$n_P(L_y,d)$. For example, for $L_y=3$, the $\lambda_{3,j}$ with coefficients
$\kappa^{(d)}$, $d=4,3,2$, and 1 are, respectively, roots of equations of
degree 1,3,5, and 4.  This case illustrates the possibility that the degree of
the corresponding equation is equal to $n_P(L_y,d)$.  A case illustrating the
possibility that it is less is, e.g., that for $L_y=4$, where the
$\lambda_{4,j}$ for $2 \le j \le 5$ with coefficients $\kappa^{(4)}$ are roots
of a cubic equation and a linear equation.

\item

Our general structural analysis shows that for a given $L_y$, there is a 
unique term $\lambda_{L_y,1}$ with coefficient of maximal degree, and this
coefficient is $\kappa^{(L_y+1)}$.  From our calculations we infer the general
result that $\lambda_{L_y,1}=(-1)^{L_y}$.  

\item 

Another inference concerns the set of terms $\lambda_{L_y,j}$ with
coefficients $\kappa^{(L_y)}$.  There are $L_y$ of these terms as shown in
(\ref{npclyminus1}). Let $\bar\lambda_{L_y,j}=(-1)^{L_y+1}\lambda_{L_y,j}$.
Then the $\bar\lambda_{L_y,j}$'s and hence the $\lambda_{L_y,j}$'s with
coefficients $\kappa^{(L_y)}$ can be calculated as follows.  Denote the
equation whose solution is $\bar\lambda_{L_y,j}$ as $f(L_y,\xi)$. Thus,
$f(1,\xi)=\xi-(q-2)$ and
\beq
f(2,\xi)=f(1,\xi)(\xi-q+3) - 1 \ .
\label{f2xi}
\eeq
The $\lambda_{L_y,j}$'s for higher values of $L_y$ are then given by
\beq
f(L_y,\xi) = f(L_y-1,\xi)(\xi-q+3)-f(L_y-2,\xi) \quad {\rm for}
\ \ L_y \ge 3 \ .
\label{flyxi}
\eeq
We find that (i) if $L_y=1$ mod 3, one of the $\lambda_{L_y,j}$'s with
coefficient $\kappa^{(L_y)}$ is $\lambda_{L_y,j}=(-1)^{L_y+1}(q-2)$; (ii) if
$L_y=2$ mod 5, two others are $(-1)^{L_y+1}$ times the roots of
$f(2,\xi)$; (iii) if $L_y=3$ mod 7, three others are $(-1)^{L_y+1}$ times
the roots of $f(3,\xi)$; and so forth.

\item

The term that is dominant in region $R_1$ is the (real) root of an equation
with degree $n_P(L_y,1)=M_{L_y}$ of maximal magnitude and has coefficient
$\kappa^{(1)}=q$.  

\item 

The locus ${\cal B}$ for the $L_x \to \infty$ limit of the family $G_D(L_y
\times L_x)$ crosses the real axis and divides the $q$ plane into various
regions.  This is in accordance with our earlier conjecture that a sufficient
(not necessary) condition for families of graphs to yield a ${\cal B}$ that
divides the $q$ plane into different regions is that it contain global
circuits, equivalent to periodic longitudinal boundary conditions for strip
graphs.  (Recall that for strip graphs with free longitudinal boundary
conditions, ${\cal B}$ does not, in general, cross the real axis or divide the
$q$ plane into regions containing this axis \cite{strip,strip2}.) 

\item 

In each case where we have an exact solution, we find that ${\cal B}$ crosses
the real axis at $q=1$ and $q=3$.

\end{itemize} 

The finding that the locus ${\cal B}$ crosses the real axis on the left at
$q=1$ is different than, but related to, our earlier finding \cite{w,wcyl},
\cite{nec}-\cite{bcc}, \cite{s4,tor4} that for the $L_x \to \infty$ limits of
lattice strip graphs with periodic longitudinal boundary conditions and either
free or transverse transverse boundary conditions, ${\cal B}$ crossed the real
axis on the left at $q=0$. For brevity of notation, we shall denote periodic
and free longitudinal boundary conditions as PBC$_x$ and FBC$_x$, respectively,
and similarly, periodic and free transverse boundary conditions as PBC$_y$ and
FBC$_y$. The present behavior can be understood as a consequence of the
property that all of the vertices on the upper side of the strip are connected
to a single external vertex.  This is especially evident for the lowest case
$L_y=1$; here, as discussed in \cite{w,wc}, it simply shifts the ${\cal B}$ for
the $L_x \to \infty$ limit of the circuit graph, $|q-1|=1$, one unit to the
right in the $q$ plane to form the locus (\ref{bwheel}).  Hence, in particular,
the crossing that would be present at $q=0$ for the circuit graph is shifted to
a crossing at $q=1$.  This suggests that the crossings that are present at
$q=1$ for all of the cases for which we have exact solutions are universal in
the same sense that the crossing on the left at $q=0$ was universal for the
strip graphs with PBC$_x$ and either FBC$_y$ or PBC$_y$.

One of the most interesting features of the present work is the property that
for all of the cases that we have studied, the point at which ${\cal B}$
crosses the real axis on the right, $q_c$, is also the same for different strip
widths $L_y$.  This is different from our earlier findings with lattice strip
graphs with PBC$_x$ and either FBC$_y$ or PBC$_y$.  There, we found that the
generic behavior was, for a given set of these boundary conditions, and a given
type of lattice, that $q_c(L_y)$ depended on $L_y$.  For example, for strips of
the square lattice with PBC$_x$ and FBC$_x$, $q_c=2$ for $L_y=1$ and for
$L_y=2$ \cite{w}, while $q_c \simeq 2.34$ for $L_y=3$ \cite{wcyl} and $q_c
\simeq 2.49$ for $L_y=4$ \cite{s4}.  From these results, we conjectured that
$q_c$ is a monotonically non-decreasing function of $L_y$ for lattice strips
with PBC$_x$ and FBC$_y$ \cite{bcc}.  In contrast, we showed that $q_c$ was not
a monotonic function of $L_y$ for lattice strips with PBC$_x$ and PBC$_y$.
For example, $q_c=2$ for $L_y=2$ (here the PBC$_y$ and FBC$_y$ results
coincide), and $q_c$ then increases to the value $q_c=3$ for $L_y=3$, but
decreases to $q_c \simeq 2.78$ for the next greater width, $L_y=4$ \cite{tor4}.
We inferred from these results that the crucial property (aside from having
PBC$_x$ so that ${\cal B}$ is guaranteed to cross the real axis and define a
$q_c$) is the type of transverse boundary conditions.  The transverse boundary
conditions for the $G_D$ family of strip graphs are of neither pure free nor
periodic type; the lower side of the strip has free boundary conditions while 
the vertices on the upper side are all connected to a single external vertex.
We can characterize the strips with PBC$_x$ and the various transverse boundary
conditions by the number of sides with free boundary conditions; this number is
2 for FBC$_y$, 1 for our present $G_D$ family, and 0 for PBC$_y$.  It is
therefore plausible that the above-mentioned conjecture that $q_c$ is a
non-decreasing function of $L_y$ for (PBC$_x$,FBC$_y$) strips also holds for
$G_D$ strips.  Accepting the validity of this conjecture, one can immediately
infer the universality of the property $q_c=3$, independent of $L_y$, as
follows.  This property $q_c=3$ holds for at least one value of $L_y$, say,
$L_y=1$.  One also knows that in the limit $L_y \to \infty$, $q_c=3$ for the
(infinite) square lattice \cite{lenard}.  Therefore, since the conjecture
states that $q_c$ cannot decrease as $L_y$ increases and since exact results
show that it has already achieved its $L_y=\infty$ value at $L_y=1$, it must
remain at this value for all larger widths $L_y$.  

As $L_y$ increases toward infinity, one would expect that, in a certain sense,
the effect of the single external vertex connected to each of the vertices on
the upper side of the strip would become negligible.  It would follow from this
reasoning that in the limit $L_y \to \infty$, the left-hand complex-conjugate
arms of ${\cal B}$ would curve around, approaching the point $q=0$, and finally
meet at this point to form a closed component on ${\cal B}$.  Our results are
consistent with this conjecture; we find that for the two largest values of
$L_y$, where these left-hand arms are present (each with its own small region),
they do approach the origin more closely for $L_y=4$ than for $L_y=3$.  

\section{Comparisons of $W(L_{\lowercase{y}},{\lowercase{q}})$ for 
different $L_{\lowercase{y}}$ }

As we have before for various infinite-length, finite lattice strips
\cite{w2d,s4,tor4}, it is of interest to compare our exact solutions for $W$ at
a given $q$, for various widths $L_y$, with the corresponding value of $W$ for
the infinite lattice, equivalent here to the limit $L_y \to \infty$.  The range
of $q$ values for which this comparison is useful is determined by the values
for which there is nonzero ground state entropy, $W > 1$ associated with the
property that the number of ways of coloring the vertices of the strip such
that no two adjacent vertices have the same color increases exponentially fast
as $W^n$ as the number of vertices $n \to \infty$ (cf. eq. \ref{w}).  Here the
limit $n \to \infty$ is taken by sending $L_x \to \infty$.  Although the
chromatic number $\chi$ alternates between 3 and 4 depending on whether $L_x$
is even or odd, one can take the limit of infinite length with $L_x$ even, and
hence this sequence of families of graphs has $\chi=3$, i.e., the
above-mentioned proper coloring can be carried out for $q \ge 3$.  We thus make
the comparison of the values of $W(q)$ for various widths with the values of
$W(sq,q)$ for the infinite square lattice for this range $q \ge 3$.  Except for
$W(sq,q=3)=(4/3)^{3/2}$ \cite{lenard}, the values $W(sq,q)$ are not known
exactly; we use the Monte Carlo calculations that were performed as part of 
\cite{ww} for this purpose.   For the comparison, as before, we 
define the ratios 
\beq
R_W(q) = \frac{W(G_D(L_y \times \infty,q))}{W(sq,q)} \ . 
\label{rw}
\eeq
The results are shown in Table \ref{wvalues}. 

\begin{table}
\caption{\footnotesize{Comparison of values of $W(G_D(L_y \times \infty),q)$
with $W(sq,q)$ for $3 \le q \le 10$.  For each value of $q$, the quantities in
the upper line are identified at the top and the quantities in the lower line
are the values of $R_W(L_y,q)$.}}
\begin{center}
\begin{tabular}{|c|c|c|c|c|c|} \hline
$q$ & $W(1,q)$ & $W(2,q)$ & $W(3,q)$ & $W(4,q)$ & $W(sq,q)$  \\ \hline 
3 &   1     & 1.27202  & 1.36138  & 1.40596    & 1.53960..   \\
  & 0.6495  & 0.8262   & 0.8842   & 0.9132     & 1           \\
4 &   2     & 2.16831  & 2.22345  & 2.25117    & 2.3370(7)   \\
  & 0.8558  &  0.9278  & 0.9514   & 0.9632     & 1           \\
5 &   3     & 3.12490  & 3.16628  & 3.18711    & 3.2510(10)  \\
  & 0.9228  & 0.9612   & 0.9739   & 0.9803     & 1           \\
6 &   4     & 4.09973  & 4.13294  & 4.14963    & 4.2003(12)  \\
  & 0.9523  & 0.9761   & 0.9840   & 0.9879     & 1           \\
7 &   5     & 5.08310  & 5.11082  & 5.12473    & 5.1669(15)  \\
  & 0.9677  & 0.9838   & 0.9891   & 0.9918     & 1           \\
8 &   6     & 6.07124  & 6.09503  & 6.10695    & 6.1431(20)  \\
  & 0.9767  & 0.9883   & 0.9922   & 0.9941     & 1           \\
9 &   7     & 7.06236  & 7.08318  & 7.09361    & 7.1254(22)  \\
  & 0.9824  & 0.9911   & 0.9941   & 0.9955     & 1           \\
10&  8      & 8.05545  & 8.07396  & 8.08323    & 8.1122(25)  \\
  & 0.9862  & 0.9930  & 0.9953    & 0.9964     & 1           \\
\hline
\end{tabular}
\end{center}
\label{wvalues}
\end{table}

\pagebreak

The property that all of the vertices on the upper surface of the strip, as
represented in Fig. \ref{strip} are connected to an external vertex has the
effect of restricting the coloring of the strip and thereby reducing the value
of $W(G_D(L_y \times \infty),q)$ relative to the value for a strip of the
square lattice with the same periodic longitudinal boundary conditions, width,
and $q$, but free transverse boundary conditions, $W(G_s(L_y \times
\infty),FBC_y,PBC_x,q)$. In \cite{w2d} a theorem was proved that for fixed $q$,
$W(G_s(L_y \times \infty),FBC_y,PBC_x,q)$ is a monotonically decreasing
function of $L_y$ which thus approaches the value of $W(sq,q)$ for the infinite
square lattice from above.  Here, because of the effect of the coloring
restriction due to the connections of all of the vertices on the upper surface
to the single external vertex, the situation is different.  Indeed, we find
numerically that (for $q > q_c$) over the range $1 \le L_y \le 4$ where we have
obtained exact solutions, $W(G_D(L_y \times \infty),q)$ is an increasing
function of $L_y$ for fixed $q$ in the range $q \ge q_c$ where .  There are two
effects depending on the strip width $L_y$ that one can identify here, and
these act in opposite directions.  The first is that as the width increases,
the restriction, per site, due to the connections of the vertices on the upper
side to the external vertex is ameliorated, since these vertices occupy a
smaller fraction of the total number of vertices as $L_y$ increases.  This is
the dominant effect and tends to increase $W$.  The second is that on the other
(lower) side of the strip there is more freedom in coloring the vertices since
they have no neighbors below them.  These lower-side vertices occupy a smaller
fraction of the total as $L_y$ increases, and this tends to decrease $W$.  It
may be recalled that the proof given in \cite{w2d} for the monotonic decrease
of $W(G_s(L_y \times \infty),q)$ for fixed $q \ge q_c$ as a function of $L_y$
for the case of free transverse boundary conditions relied upon this second
effect.  In that case, the vertices on both the upper and lower sides of the
strip, which had greater freedom of coloring because of their reduced degree
(number of neighboring vertices) occupied a commensurately smaller fraction of
the total number of vertices as the strip width increased, and this yielded the
result of the theorem.  Here, however, one has a more complicated situation in
which there are the two above-mentioned countervailing effects.

As in previous works, we also record the value of $W$ at certain other points
in region $R_1$, in particular, $q=0$ and $q=1$; these are given in Table 
\ref{wvalues2}.  Here we recall the order of
limits used in our definition of $W$, (\ref{wnoncom}). With the opposite
order of limits, one has $W(q)_{nq}=0$ for $q=0,1,$ and 2, for any $L_y$.  
\begin{table}

\caption{\footnotesize{Some additional values of values of 
$|W(G_D(L_y \times \infty),q)|$.}}
\begin{center}
\begin{tabular}{|c|c|c|c|c|} \hline
$q$&$W(1,q)$ & $W(2,q)$ & $W(3,q)$ & $W(4,q)$  \\ \hline 
0 &   2      & 2.35829  & 2.50361  & 2.58283   \\ \hline
1 &   1      & 1.61803  & 1.85334  & 1.97672   \\
\hline\hline
\end{tabular}
\end{center}
\label{wvalues2}
\end{table}

\pagebreak

\section{Conclusions}

Summarizing, in this paper we have presented a number of exact results for the
partition function $Z(G_D,q,v)$ of the $q$-state Potts model and the $T=0$
antiferromagnetic special case given by the chromatic polynomial $P(G_D,q)$, on
families of self-dual strip graphs $G_D$ of the square lattice of width $L_y$
and arbitrarily great length $L_x$ with periodic longitudinal boundary
conditions. We determined (i) the general coefficients $c_{Z,L_y,j}$ and
$c_{P,L_y,d}$ in terms of Chebyshev polynomials, (ii) the numbers $n_Z(L_y,d)$
and $n_P(L_y,d)$ of terms in $Z$ and $P$ with each type of coefficient, and
(iii) the total number of terms $N_{Z,L_y,\lambda}$ and $N_{P,L_y,\lambda}$.
We have pointed out interesting connections between the $n_Z(L_y,d)$ and
Temperley-Lieb algebras and between the $N_{F,L_y,\lambda}$, $F=Z,P$, and
enumerations of directed lattice animals.  We proceeded to present exact
calculations of $P$ for $2 \le L_y \le 4$ and to study the analytic structure
of the resultant $W$ functions in the complex $q$ plane.  A particularly
interesting finding is a universal value, $q_c=3$, for all of the widths $L_y$
considered, which, furthermore, is equal to the value for the infinite square
lattice.  This is the first family of finite-width, infinite-length lattice
strips for which we have found this type of universality of $q_c$.

{\bf Acknowledgments}

\vspace{3mm}

This research was supported in part by the U. S. NSF grant PHY-97-22101.

\vspace{4mm}

\section{Appendix}

In the text, we gave the $\lambda_{4,j}$ or the equations defining them for $1
\le j \le 5$ for the $L_y=4$ strip.  In this appendix we give the equations
for the terms $\lambda_{4,j}$ for $6 \le j \le 35$.  The $\lambda_{4,j}$, $6
\le j \le 14$, are roots of a 9-degree equation
\beqs 
& & \xi^9 -
(q^4-11q^3+51q^2-114q+101)\xi^8 \cr\cr & & -
(3q^7-51q^6+392q^5-1756q^4+4925q^3-8597q^2+8586q-3750)\xi^7 \cr\cr & & -
(3q^{10}-77q^9+899q^8-6307q^7+29495q^6-96135q^5+221090q^4-353821q^3 \cr\cr & &
+376297q^2-239411q+68889)\xi^6 \cr\cr & &
-(q^{13}-40q^{12}+701q^{11}-7302q^{10}+51090q^9-255451q^8+943367q^7-2612569q^6
\cr\cr & & +5434452q^5-8391440q^4+9348370q^3-7109828q^2+3304026q-707059)\xi^5
\cr\cr & &
+(4q^{15}-149q^{14}+2590q^{13}-27903q^{12}+208559q^{11}-1146435q^{10}
+4790250q^9
\cr\cr & & -15497328q^8+39142459q^7-77176435q^6+117779005q^5-136551251q^4
\cr\cr & & +116336276q^3-68690109q^2+25102423q-4273553)\xi^4 \cr\cr & &
-(q-2)(6q^{16}-226q^{15}+4004q^{14}-44291q^{13}+342389q^{12}-1961444q^{11}
\cr\cr & & +8613782q^{10}-29579506q^9+80262537q^8-172631424q^7+293248190q^6
\cr\cr & &
-389119576q^5+395172545q^4-296710120q^3+155196867q^2-50470243q+7677301)
\xi^3\cr\cr
& & +(q-2)(4q^{18}-167q^{17}+3293q^{16}-40754q^{15}+354781q^{14}-2307336q^{13}
\cr\cr & &
+11619216q^{12}-46328775q^{11}+148278798q^{10}-383788784q^9+805173797q^8 \cr\cr
& & -1365777711q^7+1858809772q^6-2001697759q^5+1667438591q^4 \cr\cr & &
-1036284383q^3+452215858q^2-123566922q+15898678)\xi^2 \cr\cr & &
-(q-2)^2(q^{19}-46q^{18}+994q^{17}-13427q^{16}+127231q^{15}-899290q^{14} \cr\cr
& & +4920180q^{13}-21334829q^{12}+74438607q^{11}-210900517q^{10}+487345238q^9
\cr\cr & & -918665623q^8+1407005787q^7-1735609937q^6+1699139075q^5 \cr\cr & &
-1289913357q^4+732092287q^3-292275110q^2+73185494q-8642781)\xi \cr\cr & &
-(q-2)^7(q^{15}-34q^{14}+536q^{13}-5198q^{12}+34686q^{11}-168742q^{10} \cr\cr &
& +618410q^9-1738952q^8+3783595q^7-6370774q^6+8234267q^5-8022902q^4 \cr\cr & &
+5703476q^3-2792146q^2+841318q-117542) = 0 \ .
\label{eq91}
\eeqs
The $\lambda_{4,j}$, $15 \le j \le 26$, are roots of a 12-degree equation
\beqs
& & \xi^{12} + (4q^3-33q^2+98q-103)\xi^{11} \cr\cr
& & + (6q^6-105q^5+771q^4-3057q^3+6914q^2-8454q+4350)\xi^{10} \cr\cr
& & + (4q^9-117q^8+1470q^7-10589q^6+48615q^5-148230q^4+300934q^3 \cr\cr
& & -392705q^2+298868q-100913)\xi^9 \cr\cr
& & + (q^{12}-51q^{11}+1034q^{10}-11831q^9+87645q^8-449620q^7+1652070q^6
\cr\cr
& & -4404021q^5+8481267q^4-11530332q^3+10515134q^2-5776925q+1445103)\xi^8
\cr\cr
& & -
(6q^{14}-252q^{13}+4813q^{12}-55790q^{11}+440374q^{10}-2511072q^9+10686844q^8
\cr\cr
& & -34528528q^7+85175369q^6-159716148q^5+224144839q^4-228278980q^3 \cr\cr
& & +159441815q^2-68321288q+13538233)\xi^7 \cr\cr
& & +
(15q^{16}-637q^{15}+12667q^{14}-156625q^{13}+1348190q^{12}-8567954q^{11}
\cr\cr
& & +41590251q^{10}-157306151q^9+468514489q^8-1102386000q^7+2042068500q^6
\cr\cr
& & -2946106890q^5+3244203521q^4-2634966875q^3+1487907358q^2 \cr\cr
& & -521546134q+85418708) \xi^6 \cr\cr
& & -
(20q^{18}-910q^{17}+19575q^{16}-264460q^{15}+2514176q^{14}-17858194q^{13}
\cr\cr
& &
+98204545q^{12}-427439488q^{11}+1492563991q^{10}-4211869748q^9+9626229067q^8
\cr\cr
& & -17773020220q^7+26306043996q^6-30781790215q^5+27840711649q^4 \cr\cr
& & -18773330488q^3+8883458702q^2-2630912469q+366761469)\xi^5
\cr\cr
& & +
(15q^{20}-749q^{19}+17749q^{18}-265398q^{17}+2808406q^{16}-22355292q^{15}
\cr\cr
& & +138893776q^{14}-689698353q^{13}+2779911180q^{12}-9184226013q^{11}
\cr\cr
& & +25005667770q^{10}-56201345497q^9+104079239038q^8-157931128836q^7
\cr\cr
& & +194413953469q^6-191128460116q^5+146506854315q^4-84366457483q^3 \cr\cr
& & +34322906908q^2-8792343237q+1066038269)\xi^4 \cr\cr
& &
-(q-2)(6q^{21}-324q^{20}+8279q^{19}-133226q^{18}+1515528q^{17}-12965452q^{16}
\cr\cr
& & +86626388q^{15}-463261467q^{14}+2015823729q^{13}-7215111460q^{12}
\cr\cr
& & +21384133741q^{11}-52649552494q^{10}+107694098636q^9-182459513387q^8
\cr\cr
& & +254392095608q^7-288780678892q^6+262603966158q^5-186683744725q^4
\cr\cr
& & +99924573125q^3-37861221344q^2+9048442752q-1025103882)\xi^3 \cr\cr
& &
+ (q-2)(q^{23}-65q^{22}+1965q^{21}-36996q^{20}+488991q^{19}-4842151q^{18}
\cr\cr
& & +37397267q^{17}-231374113q^{16}+1168070271q^{15}-4873964424q^{14}
\cr\cr
& &
+16957771126q^{13}-49471796934q^{12}+121369147461q^{11}-250500702268q^{10}
\cr\cr
& & +434093944035q^9-628647524990q^8+754930635327q^7-743026096643q^6
\cr\cr
& & +589226915539q^5-367162228065q^4+173056035863q^3-57980197360q^2 \cr\cr
& & +12299658707q-1241302954)\xi^2 \cr\cr
& &
+ (q-2)^2(q-3)(2q^{22}-106q^{21}+2662q^{20}-42156q^{19}+472587q^{18}
\cr\cr
& & -3990653q^{17}+26367347q^{16}-139766257q^{15}+604533952q^{14} \cr\cr
& & -2158390643q^{13}+6409094853q^{12}-15896051134q^{11}+32980281410q^{10}
\cr\cr
& & -57170276403q^9+82469188225q^8-98281609520q^7+95679352146q^6 \cr\cr
& &-74828758316q^5+45863717844q^4-21212341252q^3+6958799896q^2 \cr\cr
& & -1442606958q+142022490)\xi \cr\cr
& & + (q-2)^4(q^2-5q+5)(q^{20}-46q^{19}+999q^{18}-13621q^{17}+130785q^{16}
\cr\cr
& &
-940159q^{15}+5250952q^{14}-23336758q^{13}+83832989q^{12}-245863743q^{11}
\cr\cr
& & +591990845q^{10}-1172460575q^9+1906935986q^8-2533386002q^7 \cr\cr
& & +2722446125q^6-2330163861q^5+1551214058q^4-774015177q^3 \cr\cr
& & +272288617q^2-60200487q+6289073) = 0 \ .
\label{eq12}
\eeqs
Finally, the $\lambda_{4,j}$, $27 \le j \le 35$, are roots of another 
9-degree equation
\beqs
& & \xi^9 - (6q^2-33q+48)\xi^8 + (15q^4-168q^3+707q^2-1327q+933)\xi^7
\cr\cr
& & - (20q^6-345q^5+2450q^4-9190q^3+19218q^2-21235q+9659)\xi^6 \cr\cr
& & + (15q^8-360q^7+3693q^6-21232q^5+75012q^4-167016q^3+229025q^2 \cr\cr
& & -176806q+58742)\xi^5 \cr\cr
& & - (6q^{10}-195q^9+2732q^8-21931q^7+112366q^6-385466q^5+899022q^4
\cr\cr
& & -1410181q^3+1425238q^2-838361q+217832)\xi^4 \cr\cr
& & + (q-3)(q^{11}-45q^{10}+806q^9-7979q^8+49672q^7-206895q^6+593204q^5
\cr\cr
& & -1177217q^4+1590681q^3-1397396q^2+719432q-164543)\xi^3 \cr\cr
& & +
(q-2)(3q^{12}-108q^{11}+1726q^{10}-16260q^9+100858q^8-434873q^7+1338650q^6
\cr\cr
& & -2967964q^5+4708729q^4-5217472q^3+3834660q^2-1678782q+330985)\xi^2
\cr\cr
& & + (q-2)(q-3)(3q^{12}-93q^{11}+1295q^{10}-10716q^9+58721q^8-224620q^7
\cr\cr
& & 
+615435q^6-1217860q^5+1728820q^4-1718182q^3+1135486q^2-448175q+79896)\xi
\cr\cr 
& & + (q-2)^2(q^{13}-32q^{12}+465q^{11}-4062q^{10}+23798q^9-98757q^8
\cr\cr
& & +298728q^7-667187q^6+1100852q^5-1326203q^4+1134621q^3 \cr\cr
& & -653205q^2+226903q-35923) = 0 \ .
\label{eq92}
\eeqs

\vfill
\eject
\end{document}